\documentclass[preprint,12pt]{elsarticle}

\usepackage{amssymb}
\usepackage[version=4]{mhchem}
\usepackage{microtype}
\usepackage{url}
\journal{NIST Internal Review for submission to Matter}

\begin{document}

\begin{frontmatter}



\title{Reproducible Sorbent Materials Foundry for Carbon Capture at Scale}
\cortext[equal]{These authors contributed equally to this work}

\author[mmsd]{Austin McDannald\corref{equal}}
\ead{austin.mcdannald@nist.gov}

\ead[url]{orcid.org/0000-0002-3767-926X}

\author[mmsd]{Howie Joress\corref{equal}}

\ead{howie.joress@nist.gov}
\ead[url]{orcid.org/0000-0002-6552-2972}

\author[mmsd]{Brian DeCost\corref{equal}}

\ead[url]{orcid.org/0000-0002-3459-5888}

\author[msed]{Avery E. Baumann}
\ead[url]{orcid.org/0000-0001-8513-8049}

\author[mmsd]{A. Gilad Kusne}
\ead[url]{orcid.org/ 0000-0001-8904-2087}

\author[msed]{Kamal Choudhary}
\ead[url]{orcid.org/0000-0001-9737-8074}

\author[ncnr]{Taner Yildirim}
\ead[url]{orcid.org/0000-0002-3491-1991}

\author[csd]{Daniel W. Siderius}
\ead[url]{orcid.org/0000-0002-6260-7727}

\author[mmsd]{Winnie Wong-Ng}
\ead[url]{orcid.org/0000-0002-6938-5936}

\author[mmsd]{Andrew J. Allen}
\ead[url]{orcid.org/0000-0002-6496-8411}

\author[msed]{Christopher M. Stafford}
\ead[url]{orcid.org/0000-0002-9362-8707}

\author[mmsd]{Diana Ortiz-Montalvo}
\ead[url]{orcid.org/0000-0001-7293-4476}

\affiliation[mmsd]{organization={Material Measurement Science Division, National Institute of Standards and Technology},
            addressline={100 Bureau Dr}, 
            city={Gaithersburg},
            postcode={20899}, 
            state={MD},
            country={USA}}
\affiliation[msed]{organization={Materials Science and Engineering Division, National Institute of Standards and Technology},
            addressline={100 Bureau Dr}, 
            city={Gaithersburg},
            postcode={20899}, 
            state={MD},
            country={USA}}
            
\affiliation[csd]{organization={Chemical Sciences Division, National Institute of Standards and Technology},
            addressline={100 Bureau Dr}, 
            city={Gaithersburg},
            postcode={20899}, 
            state={MD},
            country={USA}}

\affiliation[ncnr]{organization={Center for Neutron Research, National Institute of Standards and Technology},
            addressline={100 Bureau Dr}, 
            city={Gaithersburg},
            postcode={20899}, 
            state={MD},
            country={USA}}


\begin{abstract}
We envision an autonomous sorbent materials foundry (SMF) for rapidly evaluating materials for direct air capture of carbon dioxide (\ce{CO2}), specifically targeting novel metal organic framework materials.
Our proposed SMF is hierarchical, simultaneously addressing the most critical gaps in the inter-related space of sorbent material synthesis, processing, properties, and performance.
The ability to collect these critical data streams in an agile, coordinated, and automated fashion will enable efficient end-to-end sorbent materials design through machine learning driven research framework.
\end{abstract}




\end{frontmatter}


\section{Introduction}\label{sec:sample1}

\footnote{These opinions, recommendations, findings, and conclusions do not necessarily reflect the views or policies of NIST or the United States Government.}There is an urgent and inescapable need to remove carbon from the atmosphere. 
We need to remove gigametric tons of carbon dioxide (\ce{CO2}) per year by 2050 to limit global warming to 1.5 $^\circ $C and avoid the existential threat of climate change.\cite{IPCC2022} 
While renewable energy and other emission avoidant strategies are vital, so too is the removal of \ce{CO2} from ambient air, termed direct air capture (DAC).\cite{10.17226/25259} Even if all the current and projected \ce{CO2} emissions are avoided, we will still need to increase our ability to remove \ce{CO2} from the ambient air, since there is already too much carbon in the atmosphere to avoid dangerous amounts of warming. The development of DAC offers one potential solution to that need.

DAC is an emerging technology, and the most recent DAC facility currently only removes 4,000 metric tons of \ce{CO2} per year.\cite{SmithsonianMagazinearticle,ClimteworksOrca}
Among the different DAC strategies, one promising and economically competitive method is solid sorbent DAC.
The basic concept is that a (meso)porous material such as a zeolite or metal-organic framework (MOF) is exposed to air, \ce{CO2} is adsorbed by the material, the loaded sorbent is isolated, then regenerated (heat and vacuum), releasing the \ce{CO2} for subsequent storage, and the cycle is repeated. 
However, the maturity of solid sorbent DAC is far below the current and dire demand:

\begin{itemize}
  \item There is a vast material space that has been under explored.
  \item The relationship between material properties and DAC facility performance is not well developed. This is, in part, due to the scarcity of key data:
  \begin{itemize}
    \item Mixed-gas adsorption behavior---thermodynamic and kinetic
    \item Longevity of the sorbents over adsorption/desorption cycles
    \item Structural Stability
    \item Synthesizability
  \end{itemize}
  \item Once high performing sorbents are identified, there is a need to explore synthesis routes that can support the global industrial scale needed for DAC.
\end{itemize}

Autonomous materials acceleration platforms are well suited development problems that require navigating  large and complex synthesis spaces to meet a compound set of performance requirements.  Specifically, we envision a new sorbent materials foundry (SMF) that includes fit-for-purpose structural and performance characterization to address these issues.
This foundry will autonomously explore synthesizability for novel MOF sorbents and fill in key functional properties data that is so desperately needed.
MOFs are particularly attractive as sorbents for DAC given their the wide variety of potential porous structures and chemical functionality possible in these systems.
Previous studies have found success using machine learning to propose new MOF sorbents\cite{10.1038/s42256-020-00271-1} and predict the \ce{CO2} adsorption of MOFs from their structures\cite{10.1016/j.commatsci.2022.111388}. This SMF will build upon those efforts, taking a holistic view of the characteristics of sorbents needed for DAC at scale. 
While initial efforts will focus on liquid precursor based synthesis of MOFs, one can envision including additional synthesis instrumentation modules to expand to include other synthesis processes: zeolites,\cite{10.1002/anie.201906756} amine decorated porous supports,\cite{10.1002/anie.201906756} or membranes\cite{10.1016/j.cej.2022.137047,10.1016/j.xcrp.2022.100864} for example. 
This sorbent materials foundry will consist of a hierarchical network of autonomous experimental feedback loops (FLs) designed to glean specific, high-impact knowledge:

\begin{enumerate}
  \item Synthesis and Structural Characterization (SSC). 
  \item Gas Adsorption Characterization (GAC). 
  \item Materials Performance Characterization (MPC). 
\end{enumerate}

\begin{figure}[h!tbp]
    \centering
    \includegraphics[width=0.9\textwidth]{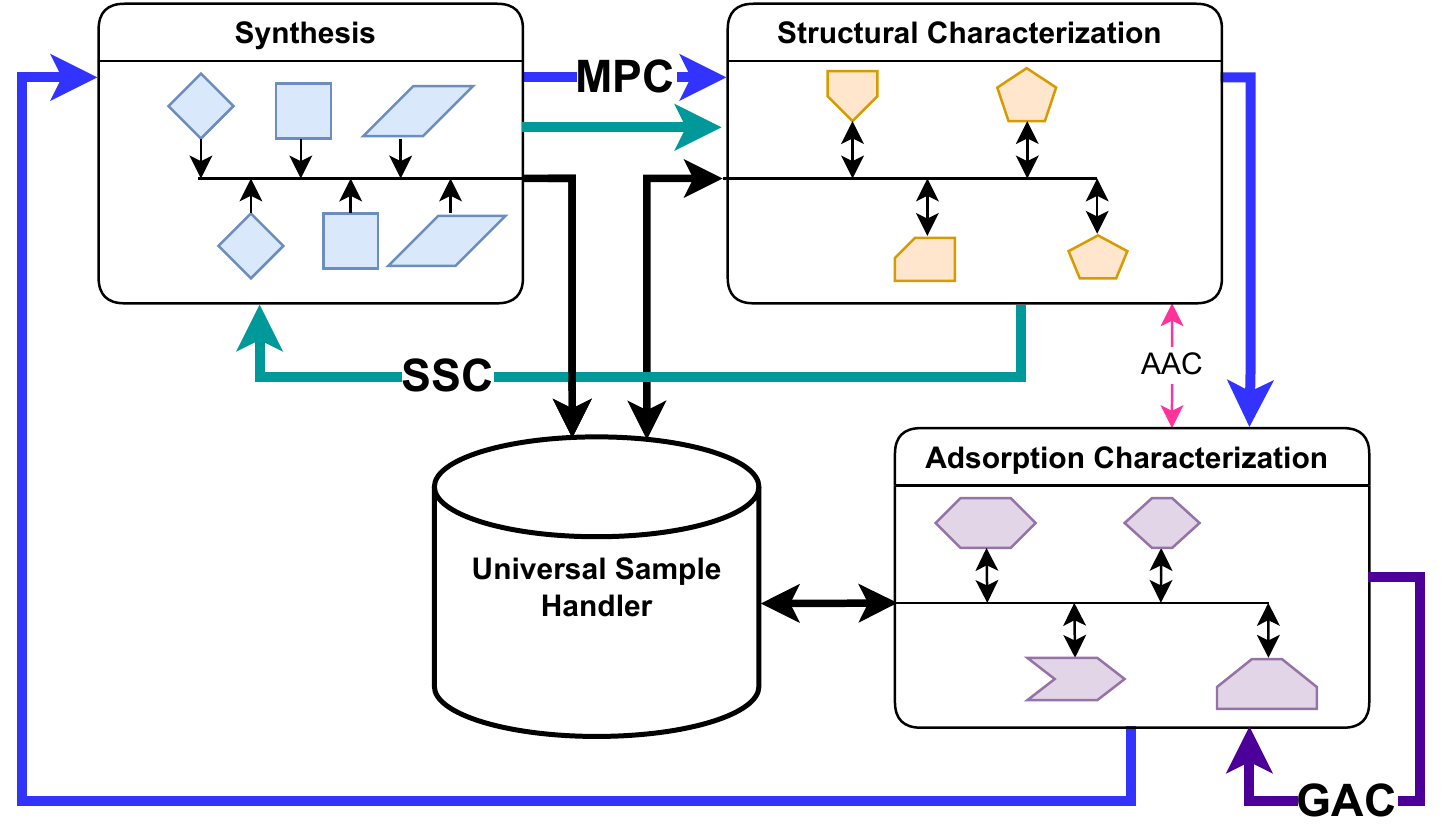}
    \caption{The Networked AI structure showing the information flow and sample movement for the autonomous feedback loops and instrument stages. The blue loop is Materials Performance Characterization (MPC). The green loop is Synthesis and Structural Characterization (SSC). The purple loop is Gas Adsorption Characterization (GAC). The pink loop is Adsorption Aging Characterization (AAC).  The black arrows represent sample movement. In the instrument clusters, the quadrilaterals are the synthesis instruments, the pentagons are the structural characterization instruments, and the hexagons are the adsorption characterization instruments.}
    \label{fig:schematic}
\end{figure}

The economy of capital research resources requires that autonomous materials research platforms are agile and modular, allowing for shifting lines of inquiry, including through addition or replacement of components.
This ensures that these tools can be used beyond the scope of a single project. 
With this in mind, the SMF will be designed to be inherently modular, consisting of a set of instrument stages connected by an automated sample handling system and controlled by the autonomous agents network structure outlined in Figure~\ref{fig:schematic}. 
Key to this modularity will be the design of a universal sample holder, capable of handling a range of solid powder samples (e.g., MOFs, Zeolites)  from the synthesis stages through each of the characterization instruments, including envisioned future stages. 
These FLs, their component instrumentation, and the engineering challenges therein are discussed in further detail below.  
We note that, while we separate these FLs for clarity and ease of discussion, in reality they will interact with each other in a coordinated autonomous campaign, increasing the overall knowledge acquisition rate.

\section{Synthesis and Structural Characterization}
\subsection{Synthesis Stage}
MOFs can be formed through a combination of metal ion clusters and organic linkers through a solvothermal or hydrothermal process. 
There are a great many of each of these two types of molecules leading to a vast number of combinations and thereby a vast number of possible structures.\cite{10.1038/nchem.1192}
Out of the space of possible MOFs, given the commonly available metal ion clusters and organic linkers, only a small percentage have ever been synthesized, and only some are stable when exposed to moisture.\cite{10.1038/nchem.1192, mofdb, 10.1002/anie.201506952, 10.1021/acs.jcim.1c00191} 
One major research thrust of this loop is to discover which of these are synthesizable and subsequently which are stable in conditions relevant to DAC applications. 
In addition to the choice of metal ion cluster and linker, other processing variables will include solvent mixture, ion and linker concentration, and temperature.  
The processing variables can have a large impact on the performance of the material. For example, longer processing times can result in larger crystals, and the crystallite size can impact the rate of adsorption. 
The synthesis stage should be flexible enough to cover a wide range of processing variables, which can then allow the elucidation of these processing-structure-property relationships.
This FL will also serve to learn and optimize these processing variables to produce high quality MOF materials. 

To achieve this experimentally, we envision a modular synthesis stage that consists of liquid precursor fluidics and powder drying/conditioning equipment. 
Since the performance characterization tools can require amounts of sample beyond the scope of pipetting robots commonly used in high-throughput and autonomous synthesis tools, a custom millifluidics system consisting of automated peristaltic and/or syringe pumps will be needed as in \cite{10.1007/s11837-022-05367-0}.  
The solution will then have to be mixed and transferred to a modular reactor for heat treatment.  While these tasks are relatively straightforward engineering tasks, with relatively low risk, the final step will be a greater engineering challenge.  
The powdered material will have to be pressed into a sample holder for subsequent characterization. Special care would be needed to ensure that sample cross-contamination is avoided. 
This will need to be done with an aim for high repeatability and designed carefully to be compatible with each characterization method.

\subsection{Structural Characterization Stage}
The first and essential feedback to the synthesis stage is the structural characterization. 
To establish whether the target material has been synthesized, the crystal structure must be determined.
 X-ray diffraction (XRD) is a versatile technique that will be used to assess the crystal structure of the synthesized sorbent powders. 
The SSC agent will use this feedback to explore possible synthesis routes and optimize recipes for each sorbent material. 
The goal of this autonomous agent is to answer questions about which materials are synthesizable by which synthesis routes.  
Automated XRD systems are readily available and further sample preparation should not be necessary for powder based methods; only minimal engineering should be required to make our sample holder compatible.

Additionally, Fourier transform infrared spectroscopy can help determine the crystal structures by the phonon mode signature.  
Crucially these methods observe how the material structure changes with exposure to moisture.
This will provide critical feedback to the autonomous agents about structural stability of these materials. Some recent work has further used this technique to (semi-quantitatively) measure the sorbent uptake.\cite{10.1021/acs.iecr.1c03756}

Microstructural characterization of the sorbent powders will also be vital. 
While this can include simple high magnification optical imaging, a key measurement here is surface area and pore size.  
This is not only important to understand how processing affects microstructure, but also for properly normalizing functional measurements.  
This type of measurement is typically performed using BET (Brunauer–Emmett–Teller) adsorption analysis.
While instrumentation for this analysis method is commercially available, there are little to no platforms designed for automated handling of samples and will therefore require customization, perhaps in collaboration with an instrument manufacturer.

The SSC agent will attempt to synthesize a given target sorbent material and determine the resultant structure and microstructure, mapping out the correlation between synthesis parameters and  structure and stability of resultant materials as well as the regions of stability in materials space. 
This knowledge is currently a critical gap for these materials, both for application of these materials to DAC as well as a variety of other critical applications, and therefore this FL will have a high impact. 

\section{Gas Adsorption Characterization}
The foundational purpose of the materials synthesized by the SMF is to assess sorbent performance for \ce{CO2} sequestration. 
Important considerations for DAC applications include  
the energy required per captured \ce{CO2},
the cycle time for the refresh cycles, 
the \ce{CO2} purity of the captured gas, 
and particularly the longevity of the sorbent performance over continued cycling. 
Each of these measures is informed from mixed gas adsorption behavior, which is imperative to predict performance in DAC relevant conditions. 
Traditionally, the mixed gas adsorption behavior has been inferred from single component isotherms or selectivity measurements at a few gas mixture conditions. 
Single component measurements are insufficient to fully characterize a sorbent’s behavior as individual analytes in mixtures interact and compete with one another, convoluting predictions of actual performance in practice.
A more holistic approach is to consider the uptake of each sorbate as a function of the partial pressure of each gas in the mixture and the temperature of the system.
The full adsorption behavior depends on the competitive adsorption of at least \ce{CO2}, \ce{N2}, \ce{O2}, and \ce{H2O} (each with an independent partial pressure) that each have their own temperature dependence.
However, mixed gas adsorption measurements are notoriously difficult and scarce,\cite{NIST_ISODB} the BISON-20 database\cite{10.4209/aaqr.2012.05.0132} of mixed gas adsorption studies only contains approximately 7,000 individual data points\cite{10.1021/acs.iecr.1c03756} and a recent study showed that $< 2\%$ of these measurements are repeatable.\cite{10.1021/acs.chemmater.7b04287}
This is particularly true in the case of DAC relevant concentrations: 450 \(\mu\)g/g (ppm) of \ce{CO2} in ambient air is an unacceptably high concentration in terms of atmospheric science but is  physicochemically challenging to measure.\cite{10.1021/acs.iecr.1c03756,10.1021/acs.iecr.9b04243,10.1016/S0001-8686(98)00048-7,10.17226/25421}
This requires careful design of the adsorption instrumentation to be able to accurately measure low concentrations of \ce{CO2}. 
An autonomous agent is well suited to fully characterize the mixed gas adsorption behavior as it is able to rapidly explore this composition+temperature-space as needed for various capture applications. 

There are several different complimentary measurement instrument designs for this mixed gas adsorption measurements.\cite{10.1021/acs.iecr.1c03756}
Open system instrument designs (commonly called breakthrough instruments) involve gas flow through a packed adsorption column where the downstream composition is measured. 
This technique has the advantages of being able to simultaneously capture both adsorption uptake and kinetic information at a specified gas composition, total pressure, and temperature. 
The disadvantages are that the measurements depend on extrinsic factors such as the packing density of the absorption bed and the head volume of the instrument design. 
The use of our envisioned automated system will ensure consistent sample preparation and aid in repeatability of the adsorption measurements.
Alternatively, there are closed systems (commonly in the volumetric or gravimetric varieties), where gas is dosed into the sorbent sample chamber. 
This has the advantage of being a more direct measurement with fewer extrinsic factors contributing variability.  
To conduct these measurements in a high-throughput manner, we propose a modular system comprising both the open and closed instrumentation that will provide complementary information and less sample-to-sample variation. 
Since the design space of independent gas mixture partial pressures and temperature is so vast, an autonomous agent will be used here to efficiently navigate this space. 
This agent will measure the adsorption at the most informative gas partial pressures and temperature conditions to fully characterize each material, distributing tasks to each type of instrument as needed. 
The adsorption measurements and their time dependance can be combined with measurements or calculations of the specific heat, heats of adsorption, structure/microstructure, and choice of refresh cycle to determine the sorbent material contributions to the performance metrics for DAC. 
Further complicating analysis, sorbents can degrade as they are cycled.
Measuring the aging of the sorbent will be accomplished by repeated adsorption and structural characterizations, as shown in the Adsorption Aging Characterization (AAC) arrows in Figure~\ref{fig:schematic}.
These will be prioritized by materials on the Pareto front of the other performance measures.

\section{Materials Performance Characterization (MPC)}
This overarching autonomous agent will combine information from both the SSC and the GAC.
This agent will then be able to address the larger scope questions. 
What sorbent materials are good for DAC? 
What high-performing materials are better suited for certain climates or cycling conditions?
The MPC agent will navigate the composition and structure space of sorbents and provide feedback on how that material can be synthesized, whether it is structurally stable, and its mixed gas adsorption behavior.
Only by accessing all this information can a holistic approach to discovering sorbents for DAC be achieved.
It is important for DAC facilities to consider the energy required to capture an amount of \ce{CO2}, the time required to cycle the sorbent, the purity of the captured \ce{CO2}, how the sorbent degrades with repeated refresh cycles, and how to obtain more sorbent. 
Each of these considerations is influenced by the sorbent material characteristics. For example, the energy per captured \ce{CO2} strongly depends on the choice of refresh cycle, which in turn is dependent on the sorbent used. 
Combining this information with information about the synthesis will allow this autonomous agent to explore the space of sorbent materials to establish the Pareto front considering their contributions to the predefined DAC performance metrics.
Once high-performing sorbents are discovered, this autonomous agent would be able to explore alternative synthesis routes to optimize processing of nano and micro structure.
This agent will combine information from the other two agents and its knowledge of chemical space of sorbents to direct the characterization measurements, thus efficiently mapping space of potential sorbents.

\section{Future Proofing}
As mentioned above, it is important for tools of this nature to be agile and retrofittable.
To this end, the modular design of this instrument allows for reuse of core components.
On the front end, synthesis stages can be added to or replaced by other approaches that produce powdered material (\textit{e.g.,} solid-state synthesis, sol-gel, spray pyrolysis) which then can be fed into our sample holder. 
Conversely, one could repurpose (or potentially use in parallel) the synthesis and structural characterization systems, resulting in a new functional characterization platform.
In particular, there is interest in MOFs for a great many important applications including catalysis, hydrogen storage, and water purification.
New stages for characterization of the performance of these materials could be designed and integrated with minimal effort.
Furthermore, due to the modular design, this system could easily adapt to emerging technologies featuring diverse material morphologies, such as gas separation membranes or catalyst architectures. 
            
\section{Resources}
Development of this SMF will require a large multidisciplinary team including MOF domain experts in material science and chemistry, chemical engineers with expertise in gas adsorption and reactor design, and mechanical engineers with expertise in automation.
In addition, computer scientists will be needed for the development of instrument controls and machine learning algorithms. We estimate this will require a dedicated team of 8 people about 3 to 5 years to complete, which includes instrumentation development and construction, and automation, algorithm development, and initial research campaigns.  
To fulfill the description of the synthesis, structural, and gas adsorption characterization stages, as well as the universal sample handler will likely require an investment of several million dollars in equipment.

\section{Conclusion}
In summary, we envision a materials acceleration platform aimed at sorbents for the direct air capture of \ce{CO2} in an effort to reduce the impact of climate change.
This would be a foundational shift in how research is performed in this field which would enable a holistic approach to characterizing adsorption behavior, heretofore absent from the literature. Furthermore, this SMF would be able to swiftly populate an otherwise sparse database of sorbent properties and performance measures.
While there are many engineering challenges for developing this system, the risks of those are mitigated by the modular design which allows the SMF to pivot to other material systems or research inquiries. 
In the short term this SMF would identify sorbent materials with the greatest suitability for DAC. 
Long term success of this SMF would allow for better data informed decision making on the design and operation of DAC facilities.


 \bibliographystyle{elsarticle-num} 
 \bibliography{cas-refs}





\end{document}